\documentclass{aa}  
\usepackage{graphicx}
\usepackage{txfonts}
\usepackage{grffile}
\usepackage{longtable}
\usepackage[font = Large]{subfig}

\def\targ{G24.78$+$0.08}

\def\nh3{NH$_{3}$}
\def\kms{km~s$^{-1}$}

\def\Vlsr{$V_{\rm LSR}$}
\def\Jyb{Jy~beam$^{-1}$}

\def\G24{G24.78$+$0.08}
\def\HII{H{\sc ii}}

\newcommand{\ms}{$M_{\odot}$}
\newcommand{\ls}{$L_{\odot}$}

\newcommand{\pas}{$\rlap{.}^{\prime\prime}$}

\newcommand{\degree}{$^{\circ}$}

\usepackage{color}

\begin{document} 

   \title{The ionized heart of a molecular disk}
   
  \subtitle{ALMA observations of the hyper-compact \HII\ region \targ\ A1}

  \titlerunning{}
   
   \author{L. Moscadelli\inst{1}
         \and
         R. Cesaroni\inst{1}
         \and
         M.T. Beltr\'an\inst{1}
         \and
         V. M. Rivilla\inst{2,1}}
  
   \institute{INAF-Osservatorio Astrofisico di Arcetri, Largo E. Fermi 5, 50125 Firenze, Italy \\
              \email{luca.moscadelli@inaf.it}
             \and
            Centro de Astrobiolog\'ia (CSIC-INTA), Ctra. de Ajalvir Km. 4, Torrej\'on de Ardoz, 28850 Madrid, Spain}
   \date{}

 
  \abstract
 {Hyper-compact (HC) or ultra-compact (UC) \HII\ regions are the first manifestations of the radiation feedback from a newly born massive star. Therefore, their study is fundamental to understanding the process of massive ($\ge 8$~\ms) star formation.}
   {We employed Atacama Large Millimeter/submillimeter Array (ALMA) 1.4~mm Cycle~6 observations to investigate at high angular resolution ($\approx$~0\pas050, corresponding to 330~au) the HC~\HII\ region inside molecular core~A1 of the high-mass star-forming cluster \targ.}
   {We used the H30$\alpha$ emission and different molecular lines of CH$_3$CN and $^{13}$CH$_3$CN  to study the kinematics of the ionized and molecular gas, respectively.}
   {At the center of the  HC~\HII\ region, at radii $\lesssim$~500~au, we observe two mutually perpendicular velocity gradients, which are directed along the axes at PA = 39\degr \ and PA = 133\degr, respectively. The velocity gradient directed along the axis at PA = 39\degr \ has an amplitude of 22~\kms~mpc$^{-1}$, which  is much larger than the other's, 3~\kms~mpc$^{-1}$. We interpret these velocity gradients as rotation around, and expansion along, the axis at PA = 39\degr. We propose a scenario where the H30$\alpha$ line traces the ionized heart of a disk-jet system that drives the formation of the massive star  ($\approx$~20~\ms) responsible for the HC~\HII\ region. Such a scenario is also supported by the position-velocity plots of the CH$_3$CN and $^{13}$CH$_3$CN lines along the axis at PA = 133\degr, which are consistent with Keplerian rotation around a 20~\ms\ star.}
   {Toward the HC~\HII\ region in \targ, the coexistence of mass infall (at radii of $\sim$~5000~au), an outer molecular disk (from $\lesssim$~4000~au\ to $\gtrsim$~500~au), and an inner ionized disk ($\lesssim$~500~au) indicates that the massive ionizing star is still actively accreting from its parental molecular core. To our knowledge, this is the first example of a molecular disk around a high-mass forming star that, while becoming internally ionized after the onset of the \HII\ region, continues to accrete mass onto the ionizing star.}


   \maketitle
%

\begin{figure*}
\centering
\includegraphics[width=\textwidth]{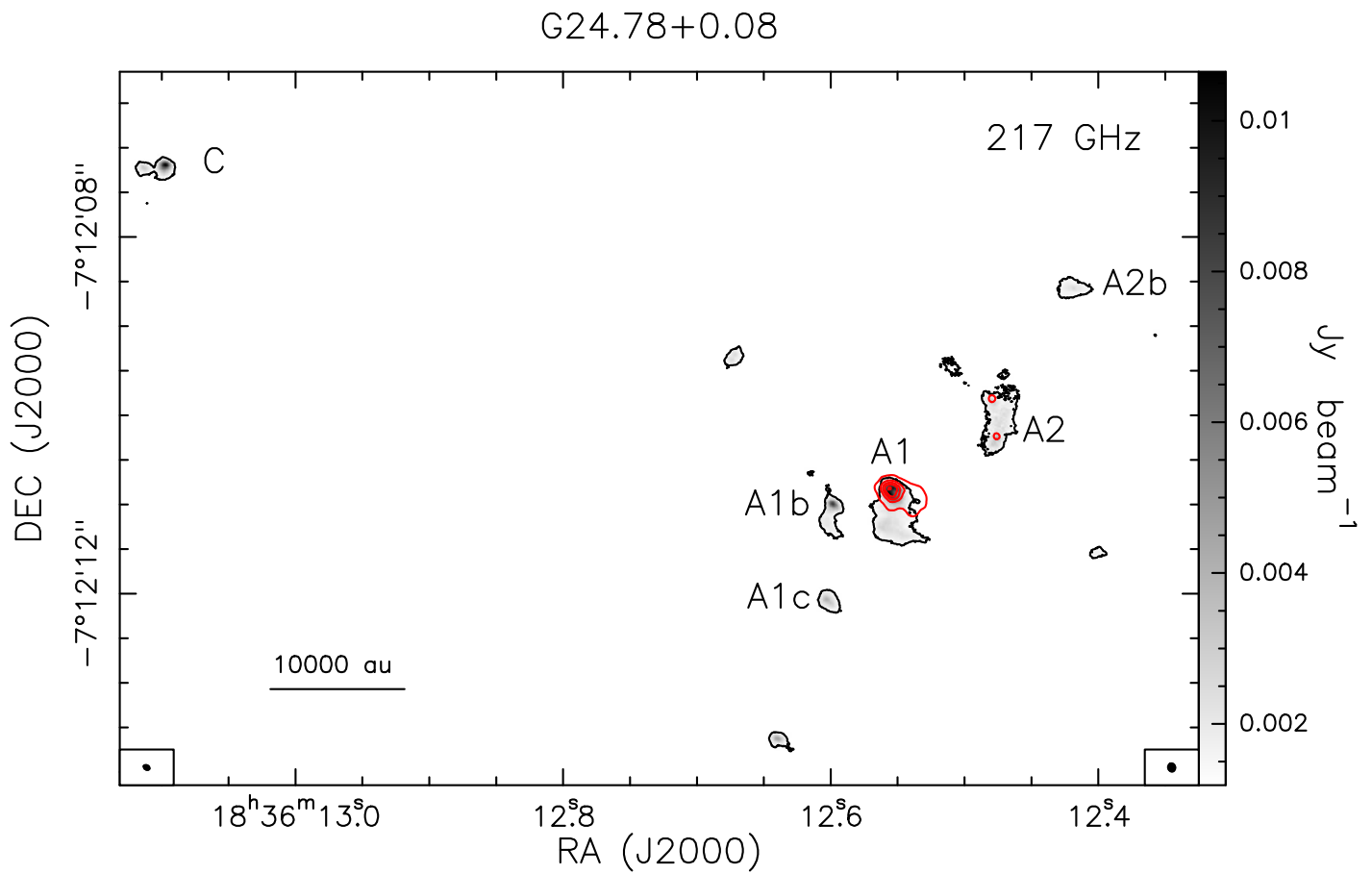} 
\caption{ALMA 2015 and JVLA observations of the high-mass star-forming region \targ. The grayscale image shows the 
ALMA 1.4~mm continuum, with the intensity scale shown to the 
right of the panel. The black contour is the 7$\sigma$ threshold of 1.2~mJy~beam$^{-1}$. The continuum emission fragments into distinct cores, labeled following \citet{Bel11}. The red contours reproduce the JVLA A-Array 1.3~cm continuum observed by \citet{Mos18}, plotting levels from 0.3 to 16.3 in steps of 4~mJy~beam$^{-1}$. The insets in the lower left and right corners of
the plot show the ALMA and JVLA beams, respectively.}
\label{G24_cont}
\end{figure*}

%
\begin{table*}
\caption{Transitions considered in this work.}             
\label{mol_trans}      
\centering                          
\begin{tabular}{c c l r}        
\hline\hline                 
Mol. Species & Frequency & Resolved QNs & $E_{\rm u}$/$k_{\rm B}$ \\    
            &  (GHz) &    &   (K) \\
\hline   
 H30$\alpha$      &  231.901  &                             &     \\
\hline   
CH$_3$CN   & 220.594  & J$_K$ =  12$_6$--11$_6$ & 326 \\
                 & 220.641  & J$_K$ =  12$_5$--11$_5$ & 247 \\
                 & 220.679  & J$_K$ =  12$_4$--11$_4$ & 183 \\
                 & 220.709  & J$_K$ =  12$_3$--11$_3$ & 133 \\
                 & 220.730  & J$_K$ =  12$_2$--11$_2$ & 97 \\
\hline                                
CH$_3$CN~v$_8$=1 & 221.299  & J$_{K,l}$ = 12$_{4,-1}$--11$_{4,-1}$ & 762 \\
                 & 221.312  & J$_{K,l}$ = 12$_{6,1}$--11$_{6,1}$   & 771 \\
                 & 221.338  & J$_{K,l}$ = 12$_{3,-1}$--11$_{3,-1}$ & 698 \\
                 & 221.350  & J$_{K,l}$ = 12$_{5,1}$--11$_{5,1}$ & 706 \\     
                 & 221.367  & J$_{K,l}$ = 12$_{2,-1}$--11$_{2,-1}$ & 649 \\
                 & 221.381  & J$_{K,l}$ = 12$_{4,1}$--11$_{4,1}$ & 655 \\
                 & 221.387  & J$_{K,l}$ = 12$_{1,-1}$--11$_{1,-1}$  & 615 \\
                 & 221.394  & J$_{K,l}$ = 12$_{0,1}$--11$_{0,1}$ & 594 \\
                 & 221.404  & J$_{K,l}$ = 12$_{3,1}$--11$_{3,1}$ & 619 \\
                 & 221.422  & J$_{K,l}$ = 12$_{2,1}$--11$_{2,1}$ & 596 \\
\hline
$^{13}$CH$_3$CN   & 232.125  & J$_K$ =  13$_5$--12$_5$ & 257 \\
                 & 232.164  & J$_K$ =  13$_4$--12$_4$ & 193 \\
                 & 232.195  & J$_K$ =  13$_3$--12$_3$ & 142 \\
                 & 232.217  & J$_K$ =  13$_2$--12$_2$ & 107 \\
                 & 232.230  & J$_K$ =  13$_1$--12$_1$ & 85 \\
                 & 232.234  & J$_K$ =  13$_0$--12$_0$ & 78 \\  
\hline
\end{tabular}
\tablefoot{\\
Column~1 reports the molecular species, Col.~2 the rest frequency of the transition, Col.~3 the quantum numbers, and Col.~4 the upper state energy.
}
\end{table*}
%

\section{Introduction}
\label{intro}
%
The process of the formation of massive ($\ge 8$~\ms \ and \ $\ge 10^4$~\ls) stars is characterized by high mass accretion rates, $\ge 10^{-4}$~\ms\ yr$^{-1}$ \citep[see, for instance,][]{Tan14}, and strong radiation feedback from the star once it reaches the zero-age main sequence (ZAMS). This feedback includes both the radiation pressure and photoionization of the circumstellar gas, which is evidenced by the appearance of a hyper-compact (HC; size $\lesssim 0.01$~pc) or ultra-compact (UC; $\lesssim 0.1$~pc) \HII\ region \citep{Kur05}. The stellar mass for the onset of an \HII\ region depends critically on the geometry and mass accretion rate onto the (proto)star \citep{Hos10}, and it is predicted to vary in the range 10--30~\ms.  Recent simulations that also account for radiation feedback \citep{Tan16,Kui18} suggest that the formation of high-mass stars, similar to that of low-mass stars,  proceeds through a disk-outflow system, in which mass accretion and ejection are intimately related, until the photoionization and radiation forces yield a progressive broadening of the cavities of the protostellar outflow and eventually quench stellar accretion completely. 

The depicted scenario can be much more complicated if the stellar mass is accreted irregularly in time, as is the case for the high-mass young stellar objects (YSOs) S255~NIRS3 \citep{Car17} and NGC6334I-MM1 \citep{Hun17}, which recently underwent a luminosity burst. In this case, as predicted by specific models \citep{Pet10} and observed for several HC~or~UC~\HII\ regions 
\citep{GalM08,FraH04,Rod07}, the \HII\ region can flicker on timescales as short as \ $\sim 10$~yr owing to rapid changes in its size, which, therefore, becomes an ambiguous indicator of age. A crucial aspect also concerns the photoevaporation of the neutral disk \citep{Hol94}, whose rate is difficult to estimate since it depends strongly on the disk geometry and (proto)stellar wind structure, which are poorly constrained by observations at the relevant small ($\sim$~100~au) scales.

In the last ten years, the high angular resolution ($\sim$ 0\farcs1) and sensitivity ($\lesssim 10$~$\mu$Jy) of the Jansky Very Large Array (JVLA) have allowed us to study the properties of the centimeter continuum emission from individual high-mass YSOs. Recent JVLA surveys targeting the earliest stages of massive star formation, from infrared dark clouds to hot molecular cores, have in most cases detected weak and compact radio emission near (within 1000~au of) the YSO, which is generally interpreted in terms of an ionized wind \citep{Mos16,Pur16,Ros16,San18,Purs21}. For several targets, the elongated and knotty structure of the continuum clearly identifies a radio jet; in a few cases, knots with a negative spectral index (indicative of synchrotron emission) are also observed \citep{San19b,Ros19}. In agreement with the results from the radio continuum, \citet{Mos19} find that the three-dimensional velocity distribution of the water masers, which trace shocked gas near the YSO, is generally consistent with the predictions for disk winds (DWs) and jets. It is remarkable that the radio luminosity of the ionized winds and jets from massive YSOs is several orders of magnitude below that expected from an optically thin \HII\ region photoionized by a ZAMS star that has a bolometric luminosity equal to that of the YSO \citep[see, for instance,][their Fig.~8]{Ang18}. Instead, the radio luminosities of the observed HC~or~UC~\HII\ regions are distributed close to the optically thin limit \citep[see, for instance, Fig.~18c of][]{Tan16}. These observations indicate that the radio winds and jets are shock-ionized and represent a stage in massive star formation prior to the onset of photoionization and the development of an \HII\ region.

For the past few decades, the kinematics of the ionized gas inside  HC~or~UC~\HII\ regions has been investigated with the Very Large Array (VLA; and, more recently, the JVLA) using hydrogen radio recombination lines (RRLs) at centimeter wavelengths. In several objects, these observations discovered large ordered motions across the ionized gas on scales of\ 100--1000~au, corresponding to infall toward, rotation around, or outflow away from the ionizing star(s) \citep{Ket02b,Sew08,Ket08b,DeP20}. The advent of the Atacama Large Millimeter/submillimeter Array (ALMA) has allowed us to observe RRLs of lower quantum numbers, which trace the gas kinematics more reliably because they are less affected by pressure broadening, and achieve much higher sensitivity, to extend the study to a larger sample of weaker \HII\ regions \citep{Klaa18,Riv20}. Moreover, ALMA permits the physical conditions and the kinematics of the molecular gas adjacent to the ionized gas to be determined in unprecedented detail. The discovery of HC~or~UC~\HII\ regions in which the ionized gas undergoes a global motion of infall or rotation suggests that mass accretion onto the star can also proceed after the onset of photoionization, in agreement with the models of ``trapped''  \HII\ regions \citep{Ket03,Ket07}.
Up to now, however, although clear cases of compact ionized rotating structures  
(size $\lesssim 100$~au) have been reported in the literature \citep{ZhaQ17,Jim20,ZhaY19}, the corresponding rotation in the molecular gas surrounding the HC~or~UC~\HII\ region on scales of $\sim$1000~au has never been observed. The coexistence of both the outer molecular and inner ionized parts of the accretion disk is expected if the massive star inside the \HII\ region is still actively accreting from its parental molecular core, as predicted by the most complete models of massive star formation \citep{Tan16,Kui18}.

The high-mass star-forming region \targ\ (bolometric luminosity of 
$\sim$$2\times10^5$~\ls\ at a distance of \ 6.7$\pm$0.7\footnote{The distance toward \targ\ has recently been determined via the trigonometric parallax measurement of the 6.7~GHz methanol masers (Moscadelli et al, in preparation).}~kpc) contains a number of molecular cores distributed over $\sim$~0.1~pc (see Fig.~\ref{G24_cont}).
Inside the most prominent molecular core, A1, VLA A-Array (from 21~cm to 7~mm) observations \citep{Bel07,Ces19b} have revealed a bright ($\sim$100~mJy at 1.3~cm) HC (size $\approx$~1000~au) \HII\ region. \citet{Bel06} mapped the core with the VLA B-Array in NH$_3$ and detected red-shifted absorption toward the HC~\HII~region, suggesting mass infall on scales of 5000~au.
\citet{Mos18} observed the HC~\HII\ region with ALMA at 1.4~mm during Cycle~2 (2015), achieving an angular resolution of $\approx$~0\farcs2 (corresponding to a linear scale of $\approx$1300~au). The analysis of the H30$\alpha$ line reveals a fast  bipolar flow in the ionized gas, covering a range of local standard of rest (LSR) velocities (\Vlsr) of \ $\approx$~60~\kms. The amplitude of the \Vlsr\ gradient, 22~\kms~mpc$^{-1}$, is one of the highest observed to date toward HC~\HII\ regions. Water and methanol masers are distributed around the HC~\HII\ region, and the three-dimensional maser velocities clearly indicate that the ionized gas is expanding at a high speed ($\ge$~200~\kms) into the surrounding molecular gas.

The present paper reports on new 1.4~mm ALMA Cycle~6 observations of the \targ\ region; the angular resolution has improved to $\approx$~0\farcs050, corresponding to $\approx$~330~au at the distance of the target. The new ALMA data allow us, for the first time, to map the velocity of the ionized gas, and, at the same time, study the kinematics of the surrounding molecular gas. The ALMA observations are described in Sect.~\ref{obs}. In Sect.~\ref{res}, we present the main observational results, which are discussed in Sect.~\ref{discu}. Our conclusions are drawn in Sect.~\ref{conclu}.

\begin{figure*}
\centering
\includegraphics[width=0.60\textwidth]{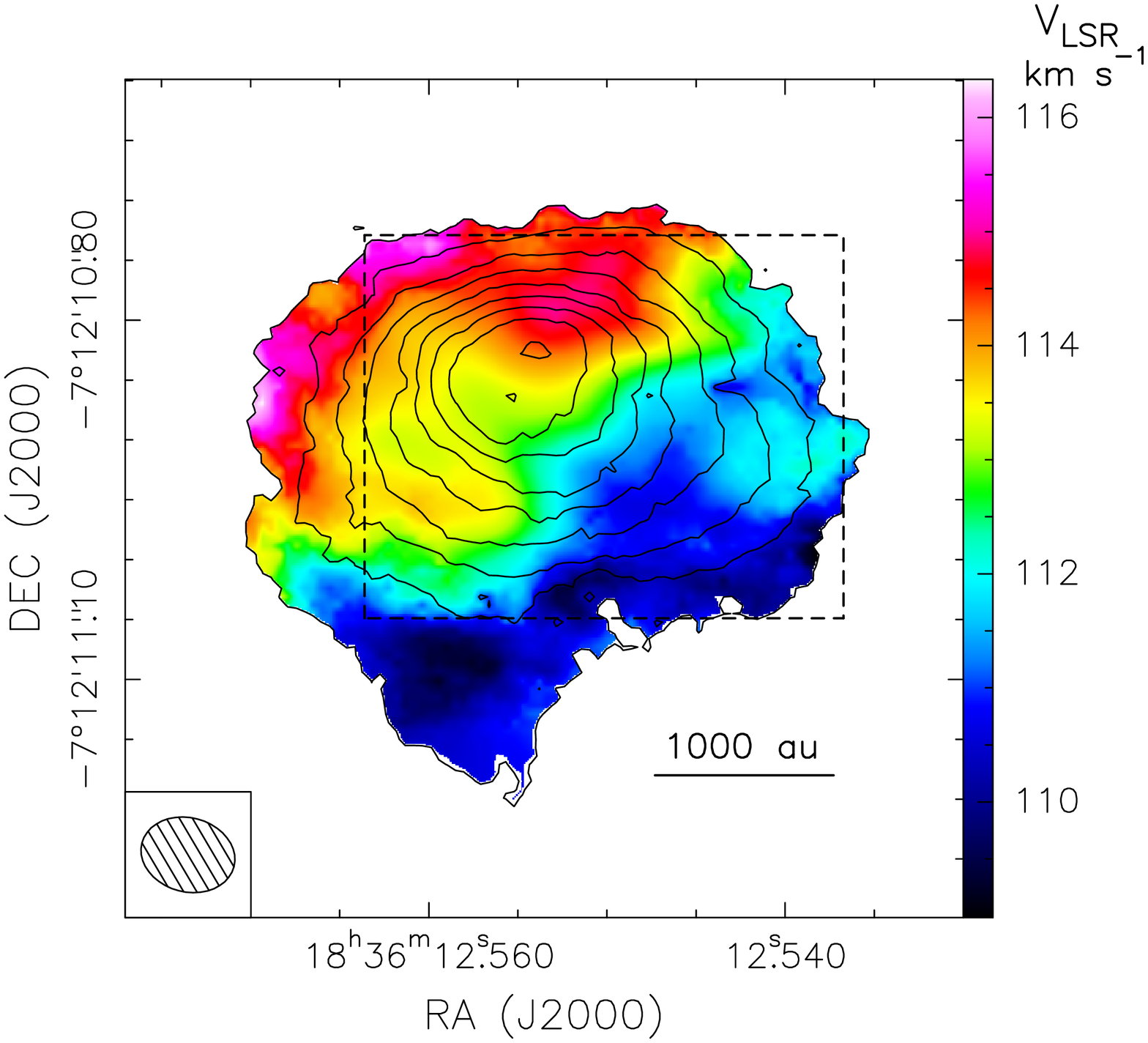}
\includegraphics[width=0.75\textwidth]{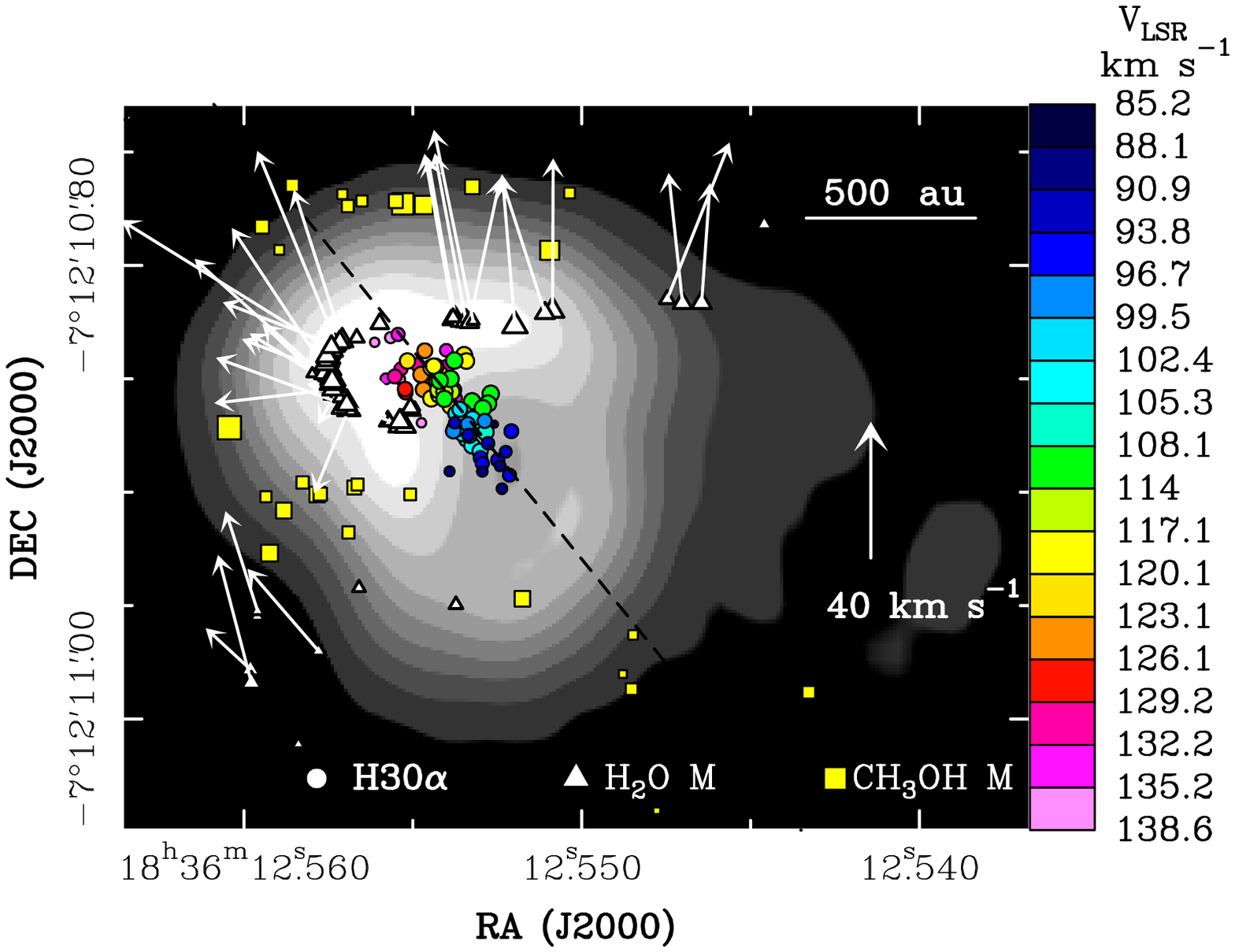}
\caption{The HC~\HII\ region. {\it Upper~panel:}~ALMA 2019 data: intensity-weighted velocity (color map) of the H30$\alpha$ emission, obtained from averaging the channel maps produced with a natural weighting of the $uv$-data. The beam (FWHM \ 0\farcs080~$\times$~0\farcs062, with \ PA $= 75$\degree) is shown in the lower~left corner. The black contours represent the velocity-integrated intensity of the H30$\alpha$ line, showing levels from \ 0.26 to 1.3 in steps of 0.13~\Jyb~\kms. The dashed black rectangle delimits the field of view shown in the lower panel. \ {\it Lower~panel:}~ALMA 2015, VLA, and Very Long Baseline Interferometry (VLBI) data. The grayscale image represents the VLA A-Array 7~mm continuum observed by \citet{Bel07}. The white triangles and yellow squares mark the VLBI positions of the H$_2$O 22~GHz \citep{Mos07} and CH$_3$OH 6.7~GHz masers \citep{Mos18}, respectively; the area of the symbol  is proportional to the logarithm of the maser intensity. The white arrows show the proper motions of the  H$_2$O masers, which were first reported and analyzed in \citet{Mos07}. The colored dots give the channel peak positions of the H30$\alpha$ line emission obtained by \citet{Mos18} from ALMA data with 0\farcs2 resolution; the colors denote the \Vlsr\ as indicated to the right of the plot. The dashed black line marks the axis of the spatial distribution of the H30$\alpha$ peaks. }
\label{Ha_nat}
\end{figure*}

\begin{figure*}
\centering
\includegraphics[width=0.75\textwidth]{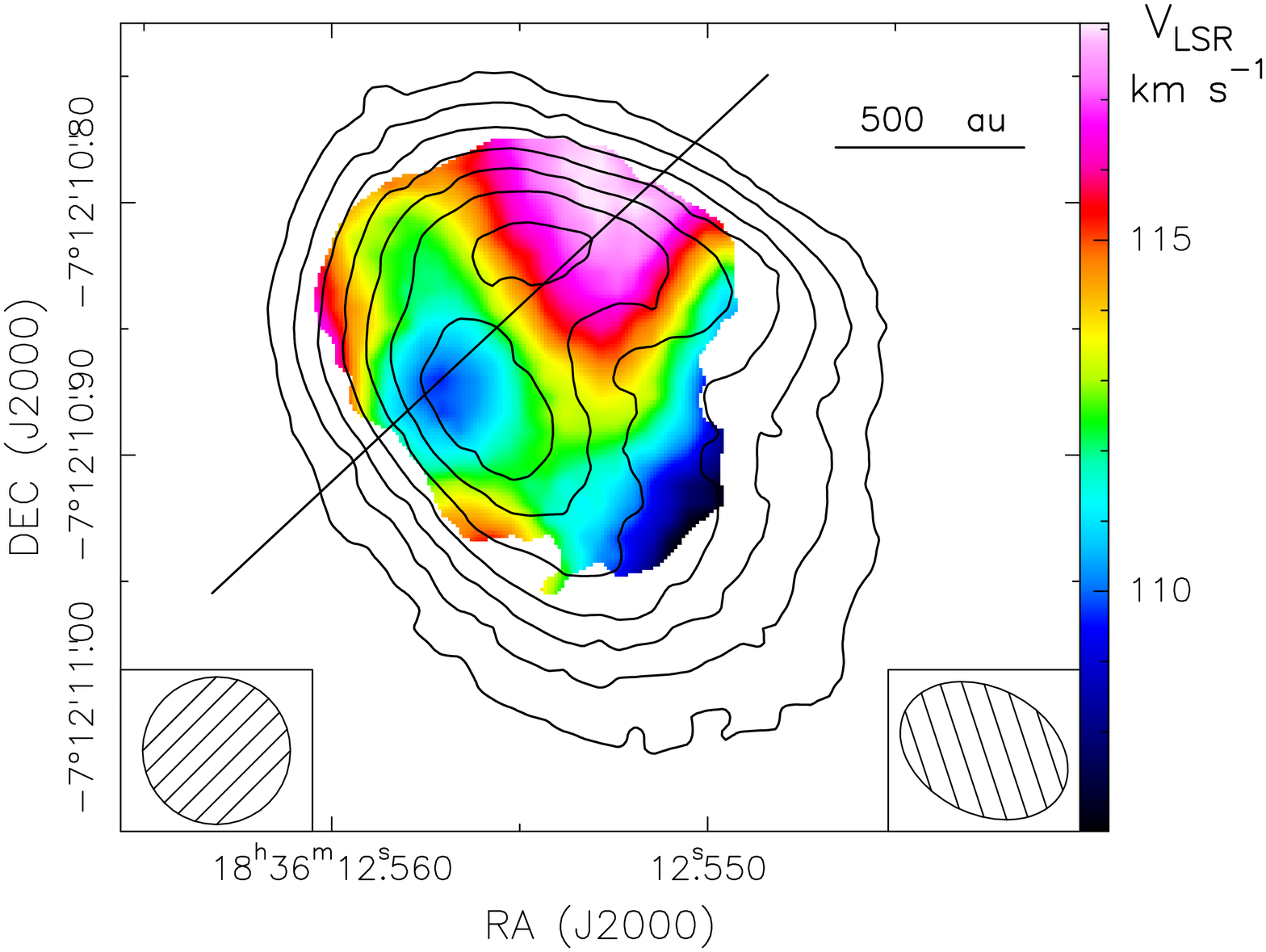}
\includegraphics[width=0.6\textwidth]{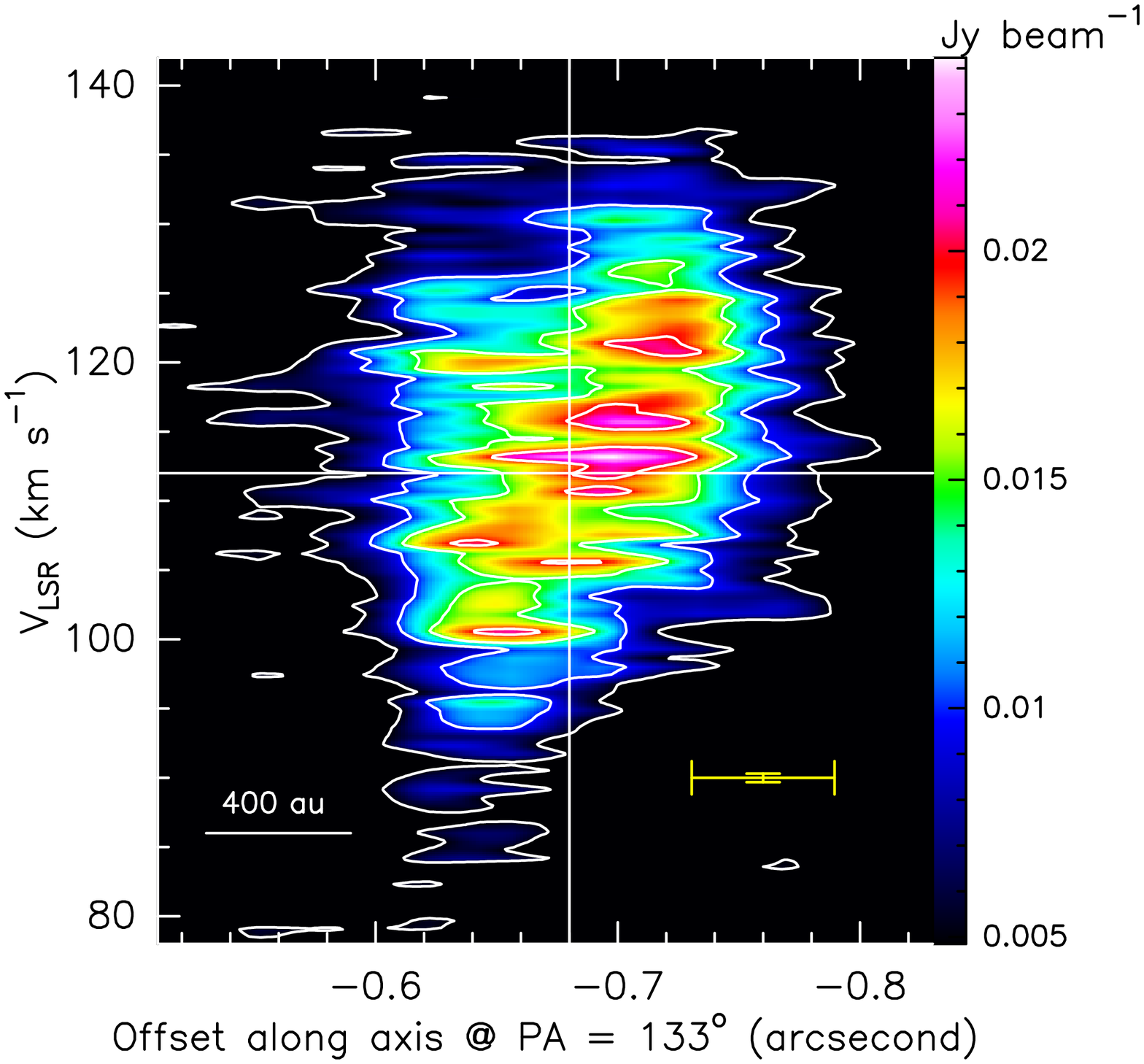}
\caption{ALMA 2019 data. {\it Upper~panel:}~Intensity-weighted velocity (color map) of the H30$\alpha$ emission, obtained from averaging the channel maps produced by weighting the $uv$-data with the ``Briggs'' robust parameter set to 0.5. The circular restoring beam has a FWHM size  (0\farcs058) equal to the geometric mean of the FWHM major and minor sizes of the Briggs beam and is shown in the lower~left corner. The straight line indicates  the direction, at PA = 133\degr, that connects the pixels of maximum and minimum velocity. The ALMA 1.4~mm continuum is represented with black contours, ranging from 3.2 to 10.6 in steps of 1.1~m\Jyb. The beam of the continuum map is reported in the lower~right corner.\ {\it Lower~panel:}~Color map and white contours showing the PV plot of the H30$\alpha$ line along the cut at PA = 133\degr.
The contour levels range from  4.9 to 24.6 in steps of 4.9~m\Jyb. The vertical and horizontal white axes denote the positional offset ($\approx$ $-$0\farcs68) and \Vlsr \ ($\approx$ 112~\kms) of the YSO, respectively. In the lower~right corner, the vertical and horizontal yellow error bars indicate the velocity and spatial resolutions, respectively}
\label{Ha_r05-ave}
\end{figure*}

\section{ALMA observations}
\label{obs}
ALMA observed \targ\ on 2019 July 29 during Cycle~6 while the 12~m array, which included  45 antennas, was in the C43-8 configuration.  The baselines ranged from  92~m to 8.5~km, for a maximum recoverable angular scale of 0\farcs8. The observations lasted $\approx$~90~min, of which $\approx$47~min were devoted to the target. The flux and bandpass calibrator was the quasar J1924$-$2914, and the phase calibrator was the quasar J1832$-$1035.

The systemic velocity of \G24, employed for Doppler correction during observations, was $V_{sys} = 111.0$~\kms. Thirteen spectral windows (SPWs) were observed over the frequency range 216.0--237~GHz: one broad (bandwidth of 1.9~GHz) SPW, for sensitive continuum measurement, and twelve narrow (0.23~GHz) SPWs, to achieve high velocity resolution (0.33--0.66~\kms) for specific spectral lines, such as the CH$_3$CN J$_K$ = 12$_{K}$--11$_{K}$ ($K = 0-6$) transitions, which are suitable for studying the kinematics and physical conditions of the gas. Data calibration was performed using the pipeline, version~``42866M'', for ALMA data analysis in the Common Astronomy Software Applications \citep[\textsc{CASA};][]{McM07} package, version~5.6. The image of each SPW  (continuum plus line emission) was produced manually using the \textsc{TCLEAN} task with the robust parameter of \citet{Bri95} set to 0.5 as a compromise between resolution and sensitivity to extended emission. The clean beams of the resulting images have full width at half maximum (FWHM) major and minor sizes in the range  \ 0\farcs070--0\farcs075 and 0\farcs049--0\farcs052, and position angles (PAs) $\approx$~62\degree. To map the emission of the ionized gas at high sensitivity, we also produced a naturally weighted image of the H30$\alpha$ line, the most intense and broadest spectral feature of SPW~5. The clean beam of the naturally weighted image has an FWHM \ of 0\farcs080~$\times$~0\farcs062 and a PA $= 75$\degree.

For each SPW, to determine the continuum level of the spectra and subtract it from the line emission, we used STATCONT \citep[\url{http://www.astro.uni-koeln.de./~sanchez/statcont};][]{Sanc17}, a statistical method for estimating the continuum level at each position of the map from the spectral distribution of the intensity at that position.
The continuum image of \G24\ has a 1$\sigma$ rms noise level of 0.2~m\Jyb, which is limited by the dynamic range. The 1$\sigma$ rms noise in a single spectral channel varies in the interval 1--2~m\Jyb, depending on the considered SPW.



\begin{figure*}
\centering
\subfloat[]{\includegraphics[width=0.49\textwidth]{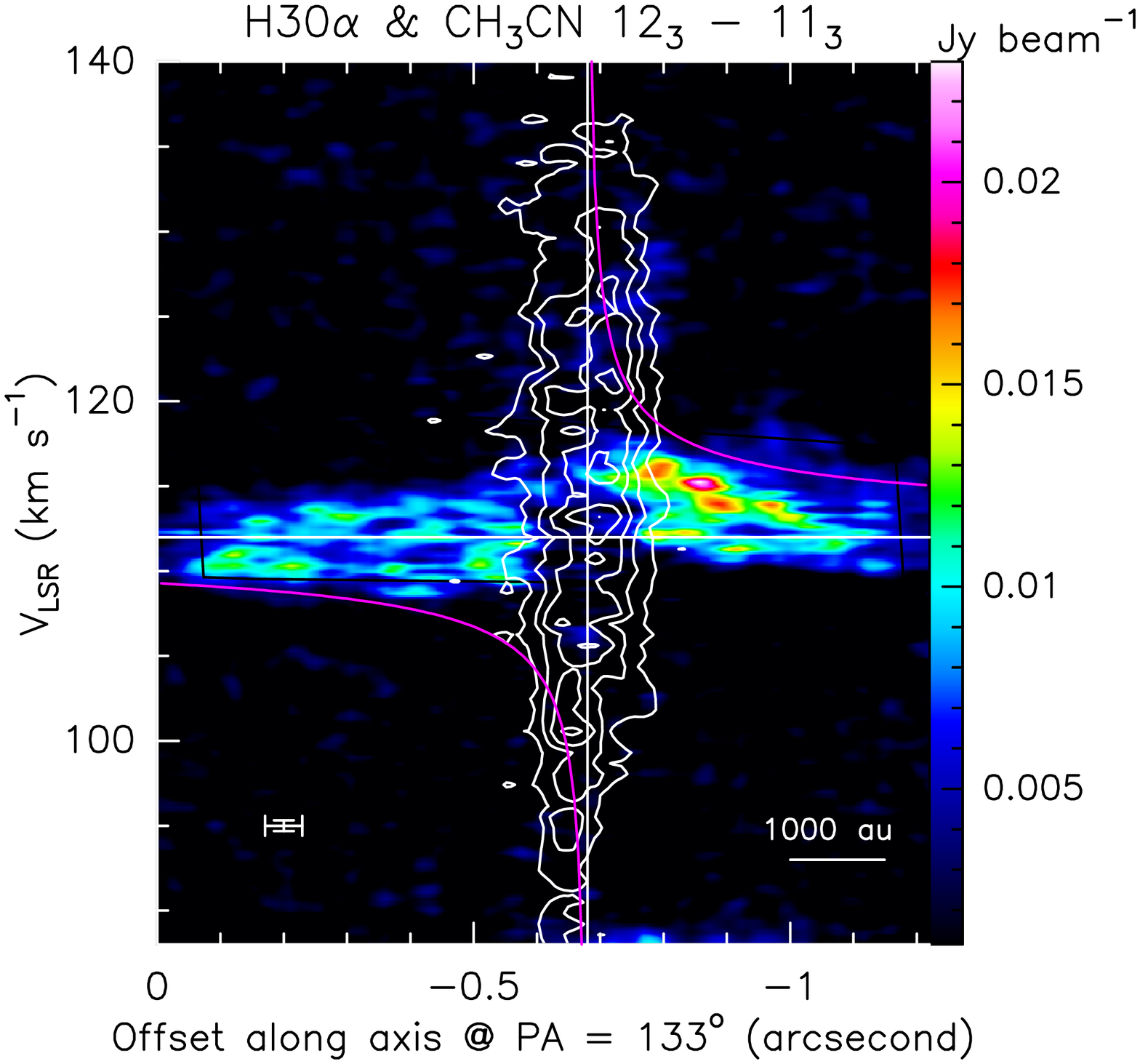}}
\subfloat[]{\includegraphics[width=0.49\textwidth]{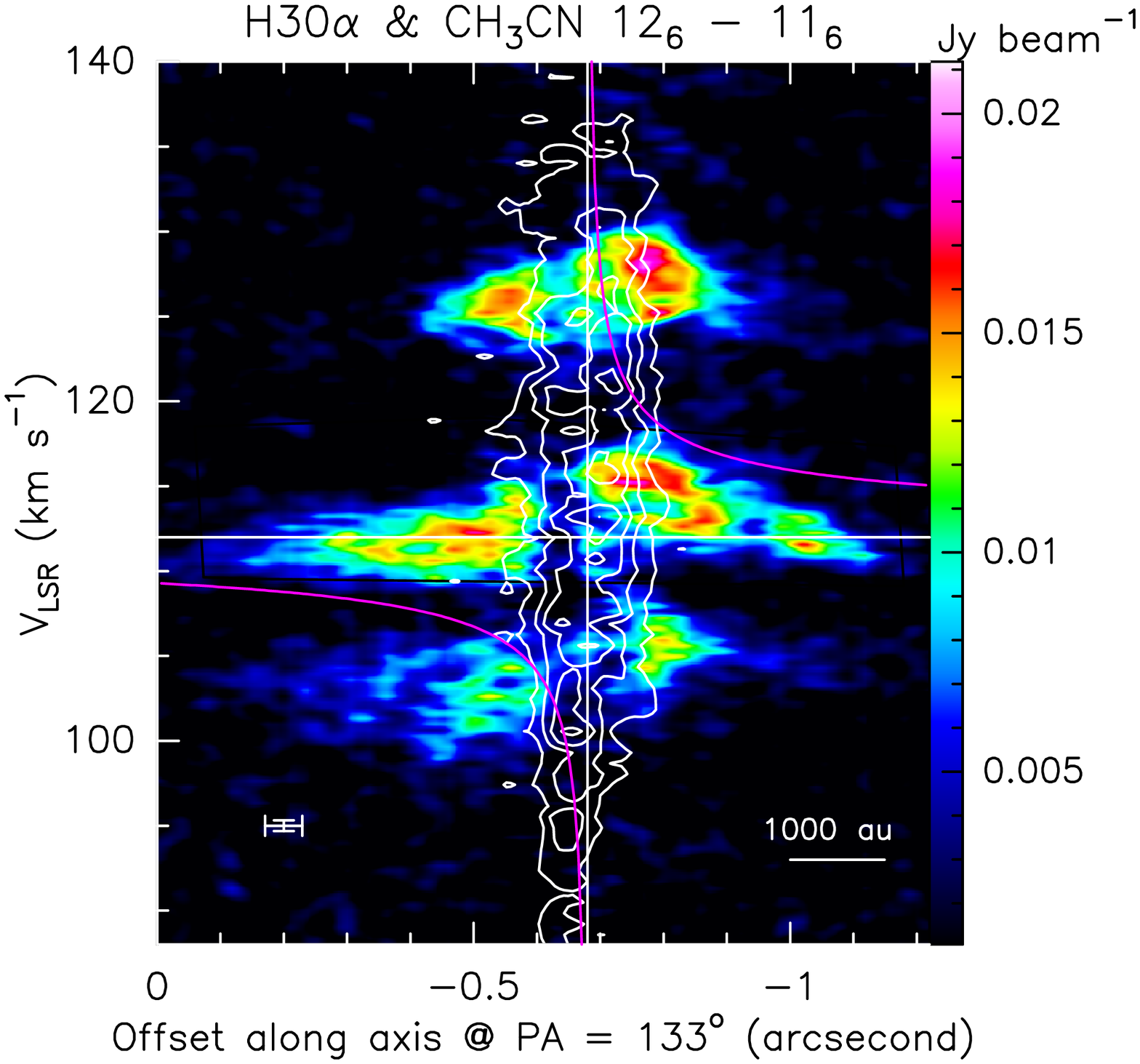}}

\subfloat[]{\includegraphics[width=0.49\textwidth]{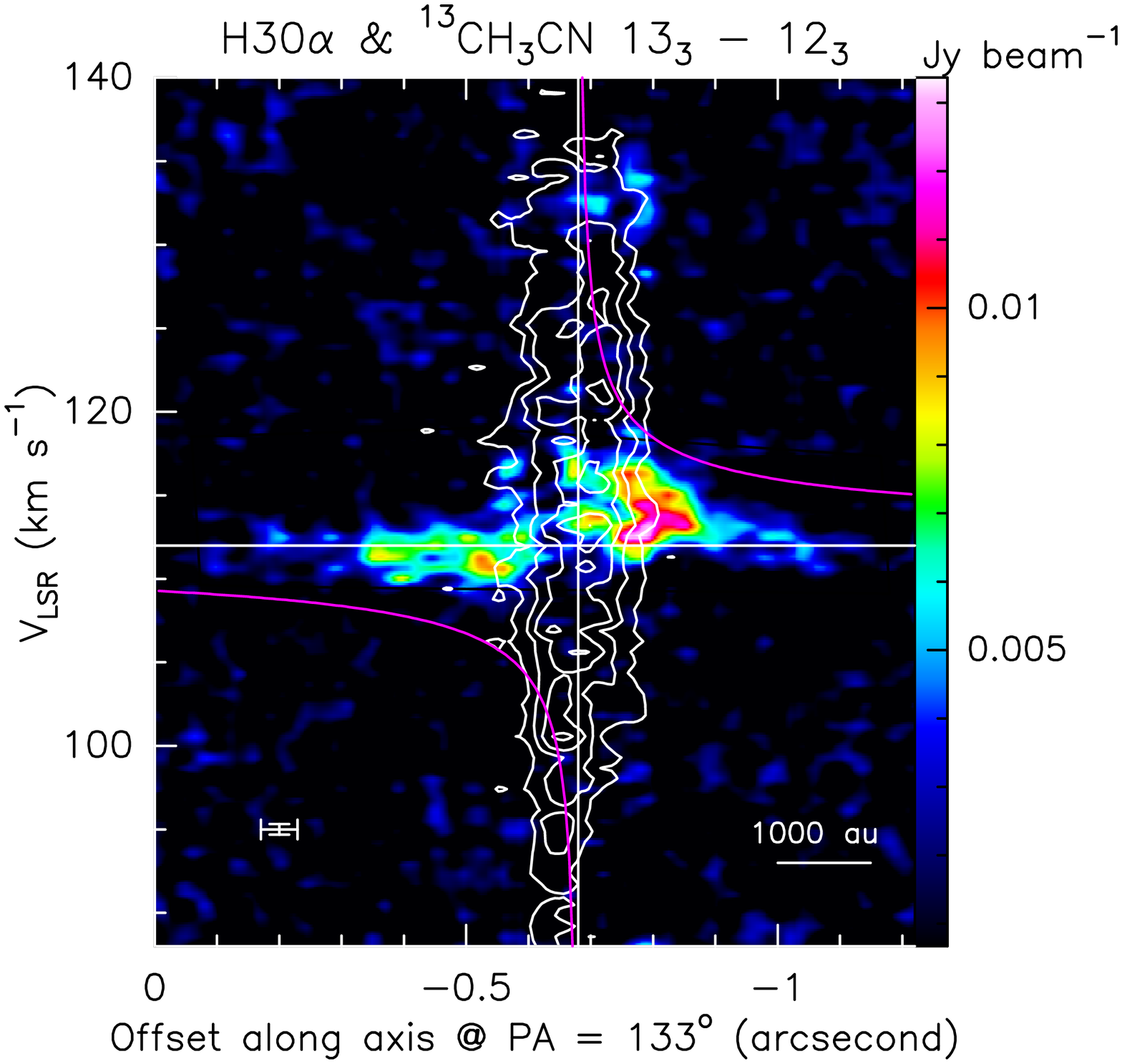}}
\subfloat[]{\includegraphics[width=0.49\textwidth]{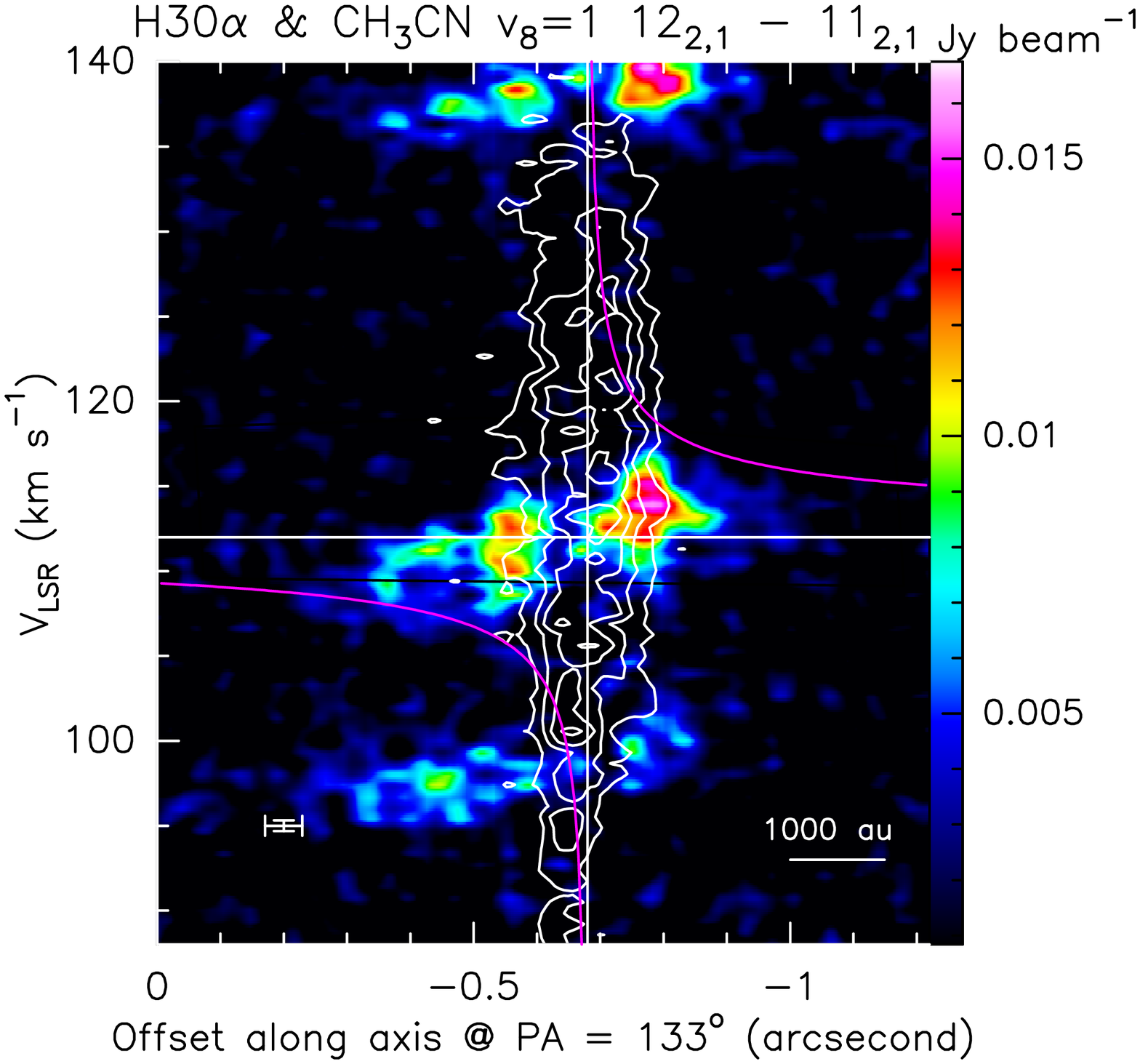}}
\caption{ALMA 2019 data. Each panel presents the combined PV plots, along the cut at PA = 133\degr, of the  H30$\alpha$ emission (white contours; same as Fig.~\ref{Ha_r05-ave}, lower~panel) and a selected molecular line (color map): CH$_3$CN J$_K$ = 12$_{3}$--11$_{3}$ (a); CH$_3$CN J$_K$ = 12$_{6}$--11$_{6}$ (b); $^{13}$CH$_3$CN J$_K$ = 13$_{3}$--12$_{3}$ (c); and CH$_3$CN~v$_8$=1 J$_{K,l}$ = 12$_{2,1}$--11$_{2,1}$ (d). Panels~(b)~and~(d) also include emission of other transitions close in frequency to the selected line. In each panel, the vertical and horizontal white axes denote the positional offset ($\approx$ $-$0\farcs68) and \Vlsr \ ($\approx$ 112~\kms) of the YSO, respectively.
The magenta curve shows the Keplerian velocity profile for a central mass of 20~\ms. In the lower~left corner, the vertical and horizontal white error bars indicate the velocity and spatial resolutions, respectively.}
\label{PV_Ha+mol}
\end{figure*}



\begin{figure*}
\centering
\includegraphics[width=\textwidth]{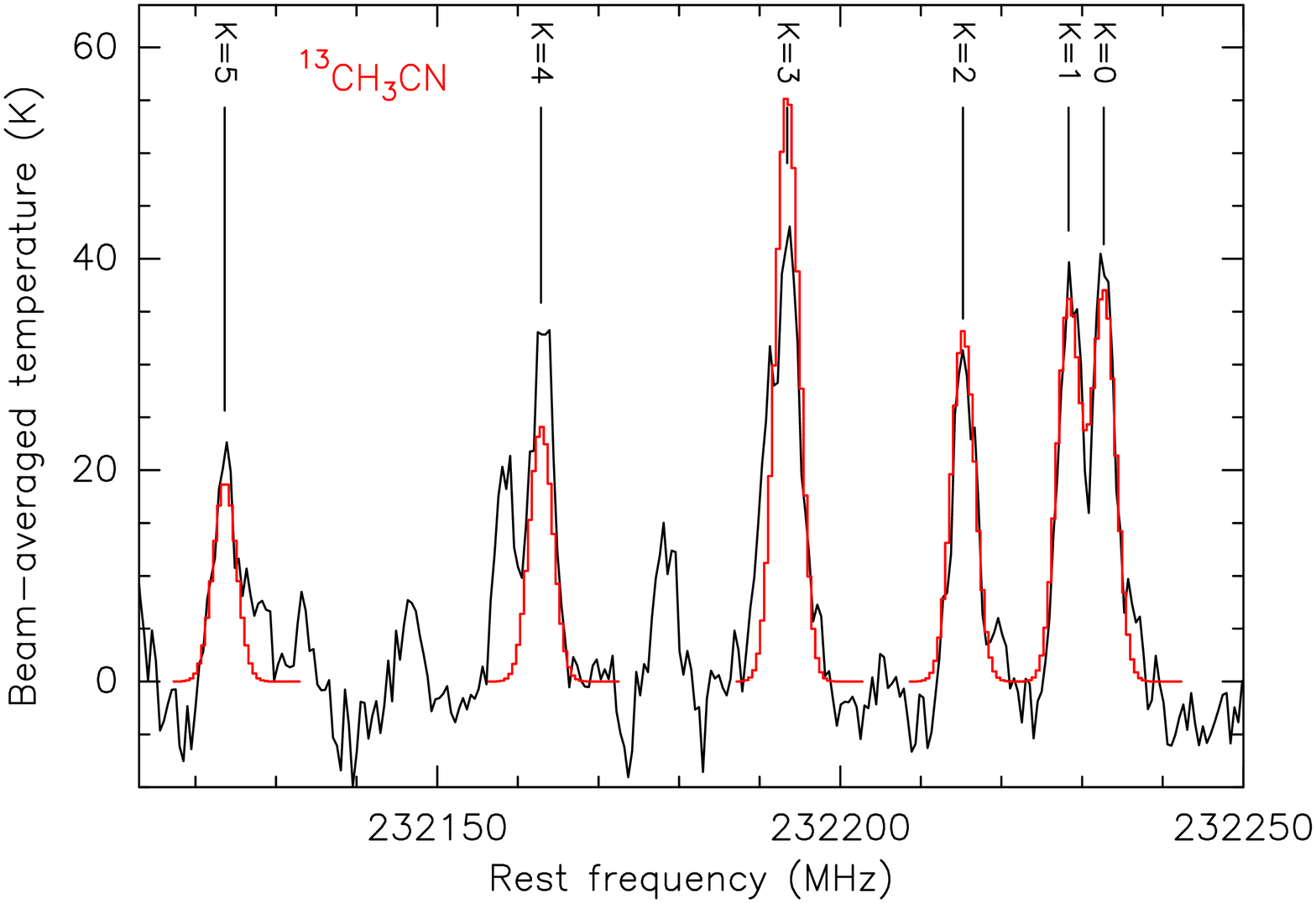}
\caption{ALMA 2019 data: observed spectrum (black line) of SPW~6 across a frequency range covering the  $^{13}$CH$_3$CN J$_K$ = 13$_K$--12$_K$, K~=~0--5, lines. The emission is averaged over a circle of 0\farcs1 radius toward the center of the HC~\HII \ region. The fit (red histogram) of the $^{13}$CH$_3$CN J$_K$ = 13$_K$--12$_K$, K~=~0--5, lines performed with \textsc{XCLASS} is also reported.}
\label{13CH3CN-fit}
\end{figure*}

\section{Results}
\label{res}

\subsection{Velocity field in the ionized gas}
\label{res-ion}
The velocity map of the ionized gas presented in Fig.~\ref{Ha_nat} (upper~panel) reveals a well-defined spatially resolved \Vlsr\ gradient, which, across a few thousand~au along the SW-NE direction, spans \ $\approx$6~\kms \ in velocity. This finding confirms the results from the previous (Cycle~2) lower-angular-resolution (beam FWHM $\approx$~0\farcs2) ALMA observations by \citet{Mos18}, who discovered a large ($\Delta$\Vlsr~$\approx$~60~\kms) SW-NE oriented (PA = 39\degr) velocity gradient in the ionized gas at the center of the HC~\HII\ region (see Fig.~\ref{Ha_nat}, lower~panel). In these previous ALMA observations, which could not spatially resolve the HC~\HII\ region, the \Vlsr\ gradient was obtained from the velocity pattern of the channel peak positions, employing a technique that provides us with relative positions as accurate as $\sim$10~mas \citep[][see their Fig.~8]{Mos18} but does not provide information about the morphology  of the emission.  With an angular resolution of \ $\approx$~70~mas, the velocity map from the new ALMA observations  (Cycle~6) unveils the spatial distribution of the slower ionized gas.  

The upper~panel of Fig.~\ref{Ha_r05-ave} shows a less sensitive but higher-angular-resolution ($\approx$~58~mas) map of the velocity field of the ionized gas. The mapped area is significantly smaller than that in  Fig.~\ref{Ha_nat} (upper~panel) and corresponds only to  the center of the HC~\HII\ region. The velocity of the ionized gas varies regularly with position and is blue-~and~red-shifted toward the SE and NW, respectively.
Taking the pixels of maximum and minimum velocities, we derive a \Vlsr\ gradient oriented at PA = 133\degr\  and with an amplitude of $\approx$~12~\kms \ across a distance of $\approx$~850~au. The lower~panel of Fig.~\ref{Ha_r05-ave} shows the position-velocity (PV) plot of the H30$\alpha$ line along the cut at PA = 133\degr.
 
\subsection{Kinematics and physical conditions of the neutral gas}
\label{res-neu}
The neutral gas surrounding the HC~\HII \ region can be traced by a plethora of molecular lines of different excitations \citep[][see their Table~2 and Figs.~2~and~3]{Mos18}; the numerous rotational transitions of CH$_3$CN and associated isotopologs are sufficiently unblended and intense to be used in our study. Table~\ref{mol_trans} lists the parameters of the lines of CH$_3$CN and $^{13}$CH$_3$CN employed in our analysis. To compare the kinematics of the molecular and ionized gas, we constructed PV plots along the direction, at PA = 133\degr\ (see Fig.~\ref{Ha_r05-ave}), of the \Vlsr\ gradient in the ionized gas detected close to the center of the HC~\HII \ region. Before producing the PV plot, the emission was averaged across three pixels (that is, 21~mas) in the direction perpendicular to the positional cut. In Fig.~\ref{PV_Ha+mol} we overlay the PV plot of the H30$\alpha$ line with that of various transitions of CH$_3$CN and  $^{13}$CH$_3$CN with upper level excitation energies ranging between 133~and~596~K.

The physical conditions of the molecular gas can be derived by fitting all the unblended lines of a given molecular species simultaneously. For this purpose,
we used the \textsc{XCLASS} (eXtended CASA Line Analysis Software Suite)
tool \citep{Moel17}. This tool models the data by solving the radiative transfer equation in one dimension for an isothermal homogeneous object in local thermodynamic equilibrium (LTE). The fitted quantities are column density, rotation temperature, velocity, line width, and source size. The fit takes the optical depth of the lines  into account as well. To determine the physical conditions of the molecular gas as close as possible to the surface of the \HII~region, we used \textsc{XCLASS} to fit the $^{13}$CH$_3$CN J$_K$ = 13$_K$--12$_K$, K~=~0--5, lines, which offer a good compromise between high signal-to-noise ratio and low-to-moderate optical depth. 
In Fig.~\ref{13CH3CN-fit} we show the fit to the spectrum obtained by averaging the $^{13}$CH$_3$CN emission over a circle of 0\farcs1 radius around the center of the \HII~region. The column density, $N_{\rm col}$, and rotational temperature, $T_{\rm rot}$, of  $^{13}$CH$_3$CN are found to be $\log_{10}\, [ N_{\rm col} \, / \, \textrm{cm$^{-2}$}] = 15.70_{-0.15}^{+0.09}$ \ and \  $T_{\rm rot} = 334_{-90}^{+80}$~K. The derived optical depths at the emission peak are 0.12, 0.12, 0.11, 0.18, 0.07, and 0.06 for the \ K = 0, 1, 2, 3, 4, and 5 lines, respectively.

\begin{figure}
\centering
\includegraphics[width=0.5\textwidth]{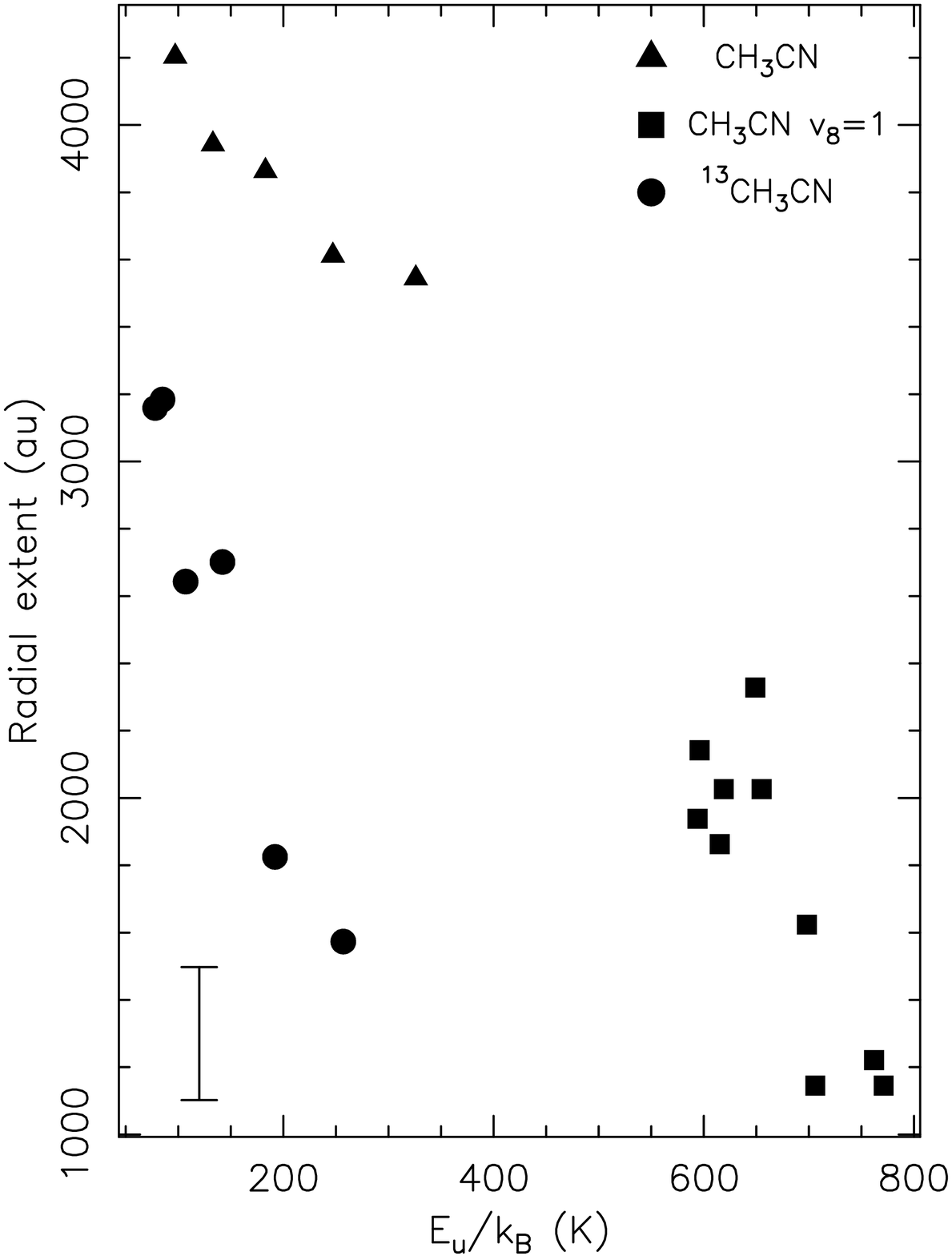}
\caption{ALMA 2019 data: plot of the radial extent of the velocity profile vs. the energy of the upper level of the transition for the molecular lines listed in Table~\ref{mol_trans}. Triangles, squares, and circles refer to transitions of CH$_3$CN, CH$_3$CN~v$_8$=1, and $^{13}$CH$_3$CN, respectively. The error bar in the lower~left corner indicates the spatial resolution.}
\label{size-excit}
\end{figure}

\section{Discussion}
\label{discu}

\subsection{Kinematics of ionized and molecular gas}
The velocity maps of the ionized gas presented in Figs.~\ref{Ha_nat}~and~\ref{Ha_r05-ave} reveal three distinct velocity components.
At the center of the  HC~\HII\ region, over $\lesssim$~500~au, we observe two mutually perpendicular \Vlsr\ gradients, directed along the axes at PA = 39\degr \ and PA = 133\degr. Remarkably, the velocity gradient directed along the axis at PA = 39\degr\ has an amplitude, 22~\kms~mpc$^{-1}$, that is much larger than that at PA = 133\degr, 3~\kms~mpc$^{-1}$.
At larger radii of a few thousand~au, we observe a relatively slow and wide-angle motion of the ionized gas along the SW-NE direction. Since the first two motions occur on similar (small) scales and are mutually perpendicular, and taking their different amplitudes into account, 
we interpret them in terms of a disk-jet system: the fast motion along the axis at PA = 39\degr \ corresponds to the ionized jet; the slower motion corresponds to the rotation of an ionized disk around the same axis. Following this interpretation, and assuming that the \Vlsr\ gradient oriented at \ PA = 133\degr\ (spanning \ $\Delta V \approx$~12~\kms \ across a distance\ $\Delta S \approx$~850~au; see Sect.~\ref{res-ion}) is due to rotation in gravito-centrifugal equilibrium, we derive a central mass\ $M_{\rm c} = ((\Delta V)^2 \ \Delta S) \ / \ (8 \ G) \approx $~17~\ms\ (where $G$ is the gravitational constant). This value is a lower limit because it does not take the disk inclination with respect to the line of sight into account and is hence consistent with the estimate of $\approx$~20~\ms \ for the mass of the ZAMS star responsible for the HC~\HII\ region \citep{Mos18}. It is plausible that the posited disk-jet system regulates the mass accretion onto the massive, ionizing star.  Moreover, if the jet is the fastest and most collimated flow portion of a DW \citep[see, for instance,][]{Pud07}, the slower and less collimated wind also naturally accounts for the slow, wide-angle expansion of the ionized gas observed at larger scales (a few thousand~au; see Fig.~\ref{Ha_nat}). 

The depicted scenario of an ionized disk-DW system inside an HC~\HII\ region is supported by recent numerical models of massive star formation, which consider the feedback of the photoionization and radiation forces from the massive star. According to \citet{Tan16}, the \HII\ region forms when the YSO has accreted between 10~and~20~\ms. While some part of the outer disk remains neutral due to shielding by the inner disk and the DW, almost the whole outflow is ionized in 10$^3$--10$^4$~yr. For low-to-moderate accretion rates \citep[][their Figs.~4~and~7]{Tan16}, the thickness of the residual flaring neutral disk around an ionizing star of  $\approx$~20~\ms\ is only a few hundred~au at a radius of 1000~au. Considering the linear resolution of $\approx$~400~au of our ALMA Cycle~6 and JVLA images of the ionized gas (see Fig.~\ref{Ha_nat}), it is not surprising that we do not detect such a thin disk buried in the ionized emission. The calculations by \citet[][see their Fig.~19]{Kui18}, which also include radiation forces, indicate that these are the dominant forces that widen the ionized outflow at scales of $\sim$1000~au and can further reduce the thickness of the neutral disk. 

Although not directly imaged, the presence of the neutral disk is recoverable from the kinematical signatures in the PV plots of the molecular tracers presented in Fig.~\ref{PV_Ha+mol}. All the tracers show an increase in the velocities going from large ($\lesssim$~4000~au) to small ($\gtrsim$~500~au) radii, which is consistent with Keplerian rotation  around a 20~\ms\ star. The molecular emission fades away close to the YSO, where the ionized gas reaches velocities much higher than those of the neutral gas. 
The dependence of the velocity profile on the excitation energy and the optical depth of the molecular line is remarkable. For each of the CH$_3$CN, CH$_3$CN~v$_8$=1, and $^{13}$CH$_3$CN transitions listed in Table~\ref{mol_trans}, we constructed the PV plot along the cut at PA = 133\degr \ and determined the radial extent of the velocity profile taking the mean of the maximum positive and negative offset (from the YSO position) for the emission at 20\% of the peak. We used a relative level to ensure that the determined size does not depend on the line intensity and chose the minimum contour at 20\% to get a signal-to-noise ratio $\gtrsim 3\sigma$, even for the weakest lines.  
Looking at Fig.~\ref{size-excit} and considering all the CH$_3$CN transitions, belonging to both the ground and excited vibrational states, it is clear that the radial extent of the velocity profile decreases regularly with the excitation energy of the line, varying from $\approx$4000~au to $\approx$1000~au  in correspondence with the increase in\ $E_{\rm u} / k_{\rm B}$ from $\approx$100~K to $\approx$800~K \citep[for qualitatively similar results toward the high-mass YSO IRAS~20126$+$4104, see Fig.14 of][]{Ces99}. 
Comparing the velocity profiles of the CH$_3$CN  and $^{13}$CH$_3$CN isotopic lines of similar excitation, it is also evident that molecules of lower optical depth trace smaller radii. 
These behaviors can have a simple explanation if, as expected, the gas temperature in the disk decreases with radius, with the consequence that lines of higher excitation, emerging from warmer gas, sample an inner portion of the disk around the YSO. A similar conclusion holds for lines of less abundant -- and thus optically thinner -- species, which trace the innermost region of the disk \citep[see also the recent results toward the O-type proto-binary system IRAS~16547$-$4247 by][]{Tana20}.

As shown in Fig.~\ref{Ha_nat} (lower~panel), the interaction of the HC~\HII\ region with the surrounding molecular environment is traced with both water and methanol masers. The latter emerge at larger separation from the ionized gas than the former, have relatively slow velocities \citep[mainly $\le$~10~\kms;][Fig.~10]{Mos18}, and trace the expansion of the ambient medium swept away by the protostellar outflow.
The lower~panel of Fig.~\ref{Ha_nat} shows that an extended arc of water masers is found just ahead of the NE lobe of the fast ionized outflow observed in the H30$\alpha$ line. Based on the fact that the water masers arise in shocks produced when the ionized flow collides against the surrounding denser molecular gas and have large proper motions ($\ge$~40~\kms,  expanding away from the center of the HC~\HII\ region; see Fig.~\ref{Ha_nat}, lower~panel),  \citet{Mos18} conclude that, at a radius of $\approx$~500~au, the velocities of the ionized flow should be $\ge$~200~\kms \ for polar angles from 0\degr\ (i.e., the flow axis) up to $\approx$45\degr. This agrees with the properties of the protostellar outflow from a 20~\ms\ star  modeled by \citet[][see their Fig.~1]{Tan16}, which, at radii $\lesssim$~1000~au, attains velocities of several hundred~\kms\ for all polar angles $\le$~45\degr.  

So far, we have commented on the good correspondence between our observations and state-of-the-art models of massive star formation. However, one critical point is the finding that, in \targ, the protostellar outflow emerging from the HC~\HII\ region has not yet expanded on a large scale but remains trapped by the surrounding dense material, as demonstrated by the water masers to the NE. This is not consistent with the simulations by \citet{Tan16}~and~\citet{Kui18}, in which, at the time of the formation of an \HII\ region, the protostellar outflow has already excavated a large cavity through the surrounding gas, up to scales $\gtrsim$~1~pc \citep[][their Fig.~4]{Kui18}.
One possible explanation of this discrepancy is that the mass accretion rate of the star ionizing the \HII\ region in \targ\ does not vary regularly with time (at variance with the above models), but instead experiences large fluctuations \citep[see, for instance, the simulations of massive star formation by][their Fig.~2, where the mass accretion rate varies by two orders of magnitude over timescales of a few hundred~years]{Mey17}. Then, because of the intermittent mass-loss rate, the outflow could never last long enough to excavate a large cavity. The outflow cavity could be periodically replenished by the higher-pressure surrounding gas during the times of minimum accretion and ejection. The fact that dense obstacles are found preferentially inside the NE lobe of the outflow, while the ionized gas freely escapes toward the SW \citep[][see their Fig.~4, upper~panel]{Mos18}, can be easily accounted for if the star is found at the near edge of the molecular core and the NE red-shifted lobe is blowing away from us toward the inner and denser region of the core.
We note that the proposed explanation in terms of a highly variable accretion rate is supported by the radiative hydrodynamical simulations of massive star formation by \citet{Pet10b,Pet10}, which successfully reproduce the short timescale variability observed in \targ\ \citep{GalM08} and other HC~or~UC~\HII\ regions \citep{FraH04,Rod07}.

\subsection{Physical conditions in the molecular disk}

We consider now the physical conditions of the gas in the molecular disk. In Sect.~\ref{res-neu}, we determined the rotational temperature,  $T_{\rm rot} = 334_{-90}^{+80}$~K, and column density, $\log_{10}\, [ N_{\rm col} \, / \, \textrm{cm$^{-2}$}]
 = 15.70_{-0.15}^{+0.09}$, of \ $^{13}$CH$_3$CN, which can be used to estimate the properties of the gas in the disk. 
Using the expression for the isotopic ratio\ $^{12}$C$/^{13}$C$~\approx 7.5 \times d_{\rm gal} + 7.6$ \citep{Wil94}, where \ $d_{\rm gal}$ \ is the galactic distance, and the value of\ $d_{\rm gal} = 3.6$~kpc suitable for the \targ \ region, we derive \ $^{12}$C$/^{13}$C~$\approx$~35. Recent detailed ALMA studies of the  CH$_3$CN abundance (with respect to H$_2$) in massive hot molecular cores indicate an average value of [CH$_3$CN/H$_2$] of $\sim$~10$^{-9}$ \citep{Pol18}. From the above values of $^{12}$C$/^{13}$C and [CH$_3$CN/H$_2$], and the $^{13}$CH$_3$CN column density, we obtain an H$_2$ column density of 
$\log_{10}\, [ N_{\rm col} \, / \, \textrm{cm$^{-2}$}] \sim 26$. 
Inspecting the PV plot of Fig.~\ref{PV_Ha+mol}c, we deduce that the $^{13}$CH$_3$CN emission is mainly found at radii between 500~au~and~1000~au. Taking 1500~au as a crude estimate of the disk diameter, we calculate the  H$_2$ number density in the disk to be \ $n_{\rm H_{2}} \sim 4 \times 10^9$~cm$^{-3}$. At such a high density, the molecular rotational transitions should be mainly excited via collisions; therefore, the rotational temperature of $^{13}$CH$_3$CN should be a reliable estimate of the gas kinetic temperature. The derived values of density, $n_{\rm H_{2}} \sim 4 \times 10^9$~cm$^{-3}$, and temperature,  
 $T_{\rm rot} = 334_{-90}^{+80}$~K, are consistent with the simulations from \citet{Tan16}, in which the molecular disk around a 20~\ms\ star, at radii $\lesssim$~1000~au, has a gas (number) density of $\sim$~10$^9$~cm$^{-3}$ \citep[][their Fig.~6, left~panel]{Tan16} and a temperature between 100~K~and~400~K \citep[][their Fig.~4]{Tan16}. 

Finally, we wanted to estimate the disk mass and the final mass of the YSO.
Assuming that the disk is close to being edge-on, its projection on the sky falls inside a rectangle centered on the HC~\HII\ region and oriented at PA = 133\degr, with a major side of 8000~au (that is, two times the radial extension of the molecular velocity profiles in Fig.~\ref{PV_Ha+mol}) and a minor side of 2000~au, which is large enough to include the whole HC~\HII\ region. Integrating the 1.4~mm continuum emission observed by \citet{Mos18} over this rectangle, we derive a flux of $\approx$~120~mJy. After correcting for the dominant free-free contribution from the HC~\HII\ region of\ $\gtrsim$ 86~mJy \citep[][see their Sect.~5.1]{Mos18}, the residual flux from dust emission is $\lesssim$~$120-86 = 34$~mJy. Assuming, as \citet{Mos18} did, a dust opacity of 1~cm$^2$~g$^{-1}$ at 1.4~mm \citep{Oss94}, a gas-to-dust mass ratio of 100, and an average dust temperature over the disk of 200~K \citep[in agreement with the aforementioned model 
by][]{Tan16}, we derive an upper limit for the total mass of molecular gas in the disk of $\lesssim$~2.5~\ms.
The mass of the ionized disk should be negligible with respect to this upper limit. We can estimate the radius of the ionized disk using the expression $ R_b = G M_\star / (2 c_{\rm s}^2)$ for the radius of a gravitationally trapped \HII\ region \citep[where  $M_\star $ is the stellar mass and $c_{\rm s}$ is the sound speed in the ionized gas;][Eq.~(3)]{Ket07}. For $c_{\rm s} = 13$~\kms\ and $M_\star = 20$~\ms, we obtain $R_b = 54$~au. For such a small radius, the density in the ionized gas should be too high, $n_{\rm H} \ge 10^{12}$~cm$^{-3}$, to  yield a non-negligible mass for the ionized disk of $\sim$1~\ms.
Therefore, we can conclude that the mass of the whole (molecular and ionized) disk is $\lesssim$~10\% of the mass, $\approx$~20~\ms, of the central star. This result is consistent with the Keplerian velocity profiles of the molecular tracers presented in Fig.~\ref{PV_Ha+mol}. 

In the scenario where the star is accreting only from its parental core,~A1, the molecular mass inside core~A1 of 4--7~\ms\ \citep[evaluated by][see their Sect.~5.1]{Mos18} represents an upper limit for the stellar mass reservoir, such that the final mass of the star should be $\lesssim$30~\ms. However, the final stellar mass could be larger than this value if the star gains its mass from the molecular envelope enshrouding A1. The elongated (SE-NW) distribution of the cores (see Fig.~\ref{G24_cont}) suggests that they are denser fragments from a more tenuous filamentary structure of molecular gas. It is therefore possible that the gravitational well of the star at the center of the HC~\HII\ region, being the most massive star in the cluster, induces  mass inflows from the nearby cores.


\section{Conclusions}
\label{conclu}

We observed the HC (size $\approx$~1000~au) \HII\ region (ionized by a ZAMS star of $\approx$~20~\ms) inside molecular core~A1 of the high-mass star-forming cluster \targ \ using ALMA at 1.4~mm during Cycle~6. By achieving an angular resolution of\ $\approx$~0\farcs050, corresponding to\ $\approx$~330~au at the target distance of 6.7~kpc, we resolved the internal kinematics of the ionized gas and found, inside a region of radius $\lesssim$~500~au, two mutually perpendicular \Vlsr\ gradients: one with a larger amplitude, 22~\kms~mpc$^{-1}$, directed along the axis at\ PA = 39\degr; the other with an amplitude of 3~\kms~mpc$^{-1}$ and oriented at\ PA = 133\degr. Along the axis at PA = 133\degr, the PV plots of different molecular tracers show similar velocity profiles that are consistent with Keplerian rotation around a 20~\ms\ star. We interpret these findings in terms of a disk that is molecular in the outer region, is internally ionized, and is in rotation around the massive star responsible for the \HII\ region, plus a fast ionized jet collimated along the disk rotation axis (at PA = 39\degr).

To our knowledge, this is the first case in which both the ionized and molecular parts of a rotating disk have been detected, extending from inside to outside an HC~\HII\ region, over radii between 100~au and 4000~au.
Toward the HC~\HII\ region in \targ, the coexistence, from large to small scales, of mass infall \citep[at a rate of $\sim$~10$^{-3}$~\ms~yr$^{-1}$ at radii of $\sim$~5000~au;][]{Bel06}, an outer molecular disk (from $\lesssim$~4000~au to $\gtrsim$~500~au), and an inner ionized disk ($\lesssim$~500~au) ensures that the massive ionizing star is still actively accreting from its parental molecular core.
These observations agree with recent models of massive star formation that include feedback of photoionization and radiation forces from the forming massive star \citep{Tan16,Kui18} and support the view of a similar disk-mediated formation process for stars of all masses.


\begin{acknowledgements}
V.M.R. has received funding from the Comunidad de Madrid through the Atracci\'on de Talento Investigador (Doctores con experiencia) Grant (COOL: Cosmic Origins Of Life; 2019-T1/TIC-15379).
\end{acknowledgements}

%
   \bibliographystyle{aa} 
   \bibliography{biblio} 

\begin{thebibliography}{50}
\expandafter\ifx\csname natexlab\endcsname\relax\def\natexlab#1{#1}\fi

\bibitem[{{Anglada} {et~al.}(2018){Anglada}, {Rodr{\'\i}guez}, \&
  {Carrasco-Gonz{\'a}lez}}]{Ang18}
{Anglada}, G., {Rodr{\'\i}guez}, L.~F., \& {Carrasco-Gonz{\'a}lez}, C. 2018,
  \aapr, 26, 3

\bibitem[{{Beltr{\'a}n} {et~al.}(2006){Beltr{\'a}n}, {Cesaroni}, {Codella},
  {Testi}, {Furuya}, \& {Olmi}}]{Bel06}
{Beltr{\'a}n}, M.~T., {Cesaroni}, R., {Codella}, C., {et~al.} 2006, \nat, 443,
  427

\bibitem[{{Beltr{\'a}n} {et~al.}(2007){Beltr{\'a}n}, {Cesaroni}, {Moscadelli},
  \& {Codella}}]{Bel07}
{Beltr{\'a}n}, M.~T., {Cesaroni}, R., {Moscadelli}, L., \& {Codella}, C. 2007,
  \aap, 471, L13

\bibitem[{{Beltr{\'a}n} {et~al.}(2011){Beltr{\'a}n}, {Cesaroni}, {Zhang},
  {Galv{\'a}n-Madrid}, {Beuther}, {Fallscheer}, {Neri}, \& {Codella}}]{Bel11}
{Beltr{\'a}n}, M.~T., {Cesaroni}, R., {Zhang}, Q., {et~al.} 2011, \aap, 532,
  A91

\bibitem[{{Briggs}(1995)}]{Bri95}
{Briggs}, D.~S. 1995, in Bulletin of the American Astronomical Society,
  Vol.~27, American Astronomical Society Meeting Abstracts, 1444

\bibitem[{{Caratti o Garatti} {et~al.}(2017){Caratti o Garatti}, {Stecklum},
  {Garcia Lopez}, {Eisl{\"o}ffel}, {Ray}, {Sanna}, {Cesaroni}, {Walmsley},
  {Oudmaijer}, {de Wit}, {Moscadelli}, {Greiner}, {Krabbe}, {Fischer}, {Klein},
  \& {Iba{\~n}ez}}]{Car17}
{Caratti o Garatti}, A., {Stecklum}, B., {Garcia Lopez}, R., {et~al.} 2017,
  Nature Physics, 13, 276

\bibitem[{{Cesaroni} {et~al.}(2019){Cesaroni}, {Beltr{\'a}n}, {Moscadelli},
  {S{\'a}nchez-Monge}, \& {Neri}}]{Ces19b}
{Cesaroni}, R., {Beltr{\'a}n}, M.~T., {Moscadelli}, L., {S{\'a}nchez-Monge},
  {\'A}., \& {Neri}, R. 2019, \aap, 624, A100

\bibitem[{{Cesaroni} {et~al.}(1999){Cesaroni}, {Felli}, {Jenness}, {Neri},
  {Olmi}, {Robberto}, {Testi}, \& {Walmsley}}]{Ces99}
{Cesaroni}, R., {Felli}, M., {Jenness}, T., {et~al.} 1999, \aap, 345, 949

\bibitem[{{De Pree} {et~al.}(2020){De Pree}, {Wilner}, {Kristensen},
  {Galv{\'a}n-Madrid}, {Goss}, {Klessen}, {Mac Low}, {Peters}, {Robinson},
  {Sloman}, \& {Rao}}]{DeP20}
{De Pree}, C.~G., {Wilner}, D.~J., {Kristensen}, L.~E., {et~al.} 2020, \aj,
  160, 234

\bibitem[{{Franco-Hern{\'a}ndez} \& {Rodr{\'\i}guez}(2004)}]{FraH04}
{Franco-Hern{\'a}ndez}, R. \& {Rodr{\'\i}guez}, L.~F. 2004, \apjl, 604, L105

\bibitem[{{Galv{\'a}n-Madrid} {et~al.}(2008){Galv{\'a}n-Madrid},
  {Rodr{\'\i}guez}, {Ho}, \& {Keto}}]{GalM08}
{Galv{\'a}n-Madrid}, R., {Rodr{\'\i}guez}, L.~F., {Ho}, P. T.~P., \& {Keto}, E.
  2008, \apjl, 674, L33

\bibitem[{{Hollenbach} {et~al.}(1994){Hollenbach}, {Johnstone}, {Lizano}, \&
  {Shu}}]{Hol94}
{Hollenbach}, D., {Johnstone}, D., {Lizano}, S., \& {Shu}, F. 1994, \apj, 428,
  654

\bibitem[{{Hosokawa} {et~al.}(2010){Hosokawa}, {Yorke}, \& {Omukai}}]{Hos10}
{Hosokawa}, T., {Yorke}, H.~W., \& {Omukai}, K. 2010, \apj, 721, 478

\bibitem[{{Hunter} {et~al.}(2017){Hunter}, {Brogan}, {MacLeod}, {Cyganowski},
  {Chandler}, {Chibueze}, {Friesen}, {Indebetouw}, {Thesner}, \&
  {Young}}]{Hun17}
{Hunter}, T.~R., {Brogan}, C.~L., {MacLeod}, G., {et~al.} 2017, \apjl, 837, L29

\bibitem[{{Jim{\'e}nez-Serra} {et~al.}(2020){Jim{\'e}nez-Serra},
  {B{\'a}ez-Rubio}, {Mart{\'\i}n-Pintado}, {Zhang}, \& {Rivilla}}]{Jim20}
{Jim{\'e}nez-Serra}, I., {B{\'a}ez-Rubio}, A., {Mart{\'\i}n-Pintado}, J.,
  {Zhang}, Q., \& {Rivilla}, V.~M. 2020, \apjl, 897, L33

\bibitem[{{Keto}(2002)}]{Ket02b}
{Keto}, E. 2002, \apj, 568, 754

\bibitem[{{Keto}(2003)}]{Ket03}
{Keto}, E. 2003, \apj, 599, 1196

\bibitem[{{Keto}(2007)}]{Ket07}
{Keto}, E. 2007, \apj, 666, 976

\bibitem[{{Keto} \& {Klaassen}(2008)}]{Ket08b}
{Keto}, E. \& {Klaassen}, P. 2008, \apjl, 678, L109

\bibitem[{{Klaassen} {et~al.}(2018){Klaassen}, {Johnston}, {Urquhart},
  {Mottram}, {Peters}, {Kuiper}, {Beuther}, {van der Tak}, \& {Goddi}}]{Klaa18}
{Klaassen}, P.~D., {Johnston}, K.~G., {Urquhart}, J.~S., {et~al.} 2018, \aap,
  611, A99

\bibitem[{{Kuiper} \& {Hosokawa}(2018)}]{Kui18}
{Kuiper}, R. \& {Hosokawa}, T. 2018, \aap, 616, A101

\bibitem[{{Kurtz}(2005)}]{Kur05}
{Kurtz}, S. 2005, in Massive Star Birth: A Crossroads of Astrophysics, ed.
  R.~{Cesaroni}, M.~{Felli}, E.~{Churchwell}, \& M.~{Walmsley}, Vol. 227,
  111--119

\bibitem[{{McMullin} {et~al.}(2007){McMullin}, {Waters}, {Schiebel}, {Young},
  \& {Golap}}]{McM07}
{McMullin}, J.~P., {Waters}, B., {Schiebel}, D., {Young}, W., \& {Golap}, K.
  2007, in Astronomical Society of the Pacific Conference Series, Vol. 376,
  Astronomical Data Analysis Software and Systems XVI, ed. R.~A. {Shaw},
  F.~{Hill}, \& D.~J. {Bell}, 127

\bibitem[{{Meyer} {et~al.}(2017){Meyer}, {Vorobyov}, {Kuiper}, \&
  {Kley}}]{Mey17}
{Meyer}, D.~M.~A., {Vorobyov}, E.~I., {Kuiper}, R., \& {Kley}, W. 2017, \mnras,
  464, L90

\bibitem[{{M{\"o}ller} {et~al.}(2017){M{\"o}ller}, {Endres}, \&
  {Schilke}}]{Moel17}
{M{\"o}ller}, T., {Endres}, C., \& {Schilke}, P. 2017, \aap, 598, A7

\bibitem[{{Moscadelli} {et~al.}(2007){Moscadelli}, {Goddi}, {Cesaroni},
  {Beltr{\'a}n}, \& {Furuya}}]{Mos07}
{Moscadelli}, L., {Goddi}, C., {Cesaroni}, R., {Beltr{\'a}n}, M.~T., \&
  {Furuya}, R.~S. 2007, \aap, 472, 867

\bibitem[{{Moscadelli} {et~al.}(2018){Moscadelli}, {Rivilla}, {Cesaroni},
  {Beltr{\'a}n}, {S{\'a}nchez-Monge}, {Schilke}, {Mottram}, {Ahmadi}, {Allen},
  {Beuther}, {Csengeri}, {Etoka}, {Galli}, {Goddi}, {Johnston}, {Klaassen},
  {Kuiper}, {Kumar}, {Maud}, {M{\"o}ller}, {Peters}, {Van der Tak}, \&
  {Vig}}]{Mos18}
{Moscadelli}, L., {Rivilla}, V.~M., {Cesaroni}, R., {et~al.} 2018, \aap, 616,
  A66

\bibitem[{{Moscadelli} {et~al.}(2016){Moscadelli}, {S{\'a}nchez-Monge},
  {Goddi}, {Li}, {Sanna}, {Cesaroni}, {Pestalozzi}, {Molinari}, \&
  {Reid}}]{Mos16}
{Moscadelli}, L., {S{\'a}nchez-Monge}, {\'A}., {Goddi}, C., {et~al.} 2016,
  \aap, 585, A71

\bibitem[{{Moscadelli} {et~al.}(2019){Moscadelli}, {Sanna}, {Cesaroni},
  {Rivilla}, {Goddi}, \& {Rygl}}]{Mos19}
{Moscadelli}, L., {Sanna}, A., {Cesaroni}, R., {et~al.} 2019, \aap, 622, A206

\bibitem[{{Ossenkopf} \& {Henning}(1994)}]{Oss94}
{Ossenkopf}, V. \& {Henning}, T. 1994, \aap, 291, 943

\bibitem[{{Peters} {et~al.}(2010{\natexlab{a}}){Peters}, {Banerjee}, {Klessen},
  {Mac Low}, {Galv{\'a}n-Madrid}, \& {Keto}}]{Pet10b}
{Peters}, T., {Banerjee}, R., {Klessen}, R.~S., {et~al.} 2010{\natexlab{a}},
  \apj, 711, 1017

\bibitem[{{Peters} {et~al.}(2010{\natexlab{b}}){Peters}, {Mac Low}, {Banerjee},
  {Klessen}, \& {Dullemond}}]{Pet10}
{Peters}, T., {Mac Low}, M.-M., {Banerjee}, R., {Klessen}, R.~S., \&
  {Dullemond}, C.~P. 2010{\natexlab{b}}, \apj, 719, 831

\bibitem[{{Pols} {et~al.}(2018){Pols}, {Schw{\"o}rer}, {Schilke}, {Schmiedeke},
  {S{\'a}nchez-Monge}, \& {M{\"o}ller}}]{Pol18}
{Pols}, S., {Schw{\"o}rer}, A., {Schilke}, P., {et~al.} 2018, \aap, 614, A123

\bibitem[{{Pudritz} {et~al.}(2007){Pudritz}, {Ouyed}, {Fendt}, \&
  {Brandenburg}}]{Pud07}
{Pudritz}, R.~E., {Ouyed}, R., {Fendt}, C., \& {Brandenburg}, A. 2007, in
  Protostars and Planets V, ed. B.~{Reipurth}, D.~{Jewitt}, \& K.~{Keil}, 277

\bibitem[{{Purser} {et~al.}(2021){Purser}, {Lumsden}, {Hoare}, \&
  {Kurtz}}]{Purs21}
{Purser}, S.~J.~D., {Lumsden}, S.~L., {Hoare}, M.~G., \& {Kurtz}, S. 2021,
  \mnras

\bibitem[{{Purser} {et~al.}(2016){Purser}, {Lumsden}, {Hoare}, {Urquhart},
  {Cunningham}, {Purcell}, {Brooks}, {Garay}, {G{\'u}zman}, \&
  {Voronkov}}]{Pur16}
{Purser}, S.~J.~D., {Lumsden}, S.~L., {Hoare}, M.~G., {et~al.} 2016, \mnras,
  460, 1039

\bibitem[{{Rivera-Soto} {et~al.}(2020){Rivera-Soto}, {Galv{\'a}n-Madrid},
  {Ginsburg}, \& {Kurtz}}]{Riv20}
{Rivera-Soto}, R., {Galv{\'a}n-Madrid}, R., {Ginsburg}, A., \& {Kurtz}, S.
  2020, \apj, 899, 94

\bibitem[{{Rodr{\'\i}guez} {et~al.}(2007){Rodr{\'\i}guez}, {G{\'o}mez}, \&
  {Tafoya}}]{Rod07}
{Rodr{\'\i}guez}, L.~F., {G{\'o}mez}, Y., \& {Tafoya}, D. 2007, \apj, 663, 1083

\bibitem[{{Rosero} {et~al.}(2016){Rosero}, {Hofner}, {Claussen}, {Kurtz},
  {Cesaroni}, {Araya}, {Carrasco-Gonz{\'a}lez}, {Rodr{\'{\i}}guez}, {Menten},
  {Wyrowski}, {Loinard}, \& {Ellingsen}}]{Ros16}
{Rosero}, V., {Hofner}, P., {Claussen}, M., {et~al.} 2016, \apjs, 227, 25

\bibitem[{{Rosero} {et~al.}(2019){Rosero}, {Hofner}, {Kurtz}, {Cesaroni},
  {Carrasco-Gonz{\'a}lez}, {Araya}, {Rodr{\'\i}guez}, {Menten}, {Wyrowski},
  {Loinard}, {Ellingsen}, \& {Molinari}}]{Ros19}
{Rosero}, V., {Hofner}, P., {Kurtz}, S., {et~al.} 2019, \apj, 880, 99

\bibitem[{{Sanchez-Monge} {et~al.}(2017){Sanchez-Monge}, {Schilke}, {Ginsburg},
  {Cesaroni}, \& {Schmiedeke}}]{Sanc17}
{Sanchez-Monge}, A., {Schilke}, P., {Ginsburg}, A., {Cesaroni}, R., \&
  {Schmiedeke}, A. 2017, ArXiv e-prints

\bibitem[{{Sanna} {et~al.}(2019){Sanna}, {Moscadelli}, {Goddi}, {Beltr{\'a}n},
  {Brogan}, {Caratti o Garatti}, {Carrasco-Gonz{\'a}lez}, {Hunter}, {Massi}, \&
  {Padovani}}]{San19b}
{Sanna}, A., {Moscadelli}, L., {Goddi}, C., {et~al.} 2019, \aap, 623, L3

\bibitem[{{Sanna} {et~al.}(2018){Sanna}, {Moscadelli}, {Goddi}, {Krishnan}, \&
  {Massi}}]{San18}
{Sanna}, A., {Moscadelli}, L., {Goddi}, C., {Krishnan}, V., \& {Massi}, F.
  2018, \aap, 619, A107

\bibitem[{{Sewi{\l}o} {et~al.}(2008){Sewi{\l}o}, {Churchwell}, {Kurtz}, {Goss},
  \& {Hofner}}]{Sew08}
{Sewi{\l}o}, M., {Churchwell}, E., {Kurtz}, S., {Goss}, W.~M., \& {Hofner}, P.
  2008, \apj, 681, 350

\bibitem[{{Tan} {et~al.}(2014){Tan}, {Beltr{\'a}n}, {Caselli}, {Fontani},
  {Fuente}, {Krumholz}, {McKee}, \& {Stolte}}]{Tan14}
{Tan}, J.~C., {Beltr{\'a}n}, M.~T., {Caselli}, P., {et~al.} 2014, Protostars
  and Planets VI, 149

\bibitem[{{Tanaka} {et~al.}(2016){Tanaka}, {Tan}, \& {Zhang}}]{Tan16}
{Tanaka}, K. E.~I., {Tan}, J.~C., \& {Zhang}, Y. 2016, \apj, 818, 52

\bibitem[{{Tanaka} {et~al.}(2020){Tanaka}, {Zhang}, {Hirota}, {Sakai},
  {Motogi}, {Tomida}, {Tan}, {Rosero}, {Higuchi}, {Ohashi}, {Liu}, \&
  {Sugiyama}}]{Tana20}
{Tanaka}, K. E.~I., {Zhang}, Y., {Hirota}, T., {et~al.} 2020, \apjl, 900, L2

\bibitem[{{Wilson} \& {Rood}(1994)}]{Wil94}
{Wilson}, T.~L. \& {Rood}, R. 1994, \araa, 32, 191

\bibitem[{{Zhang} {et~al.}(2017){Zhang}, {Claus}, {Watson}, \&
  {Moran}}]{ZhaQ17}
{Zhang}, Q., {Claus}, B., {Watson}, L., \& {Moran}, J. 2017, \apj, 837, 53

\bibitem[{{Zhang} {et~al.}(2019){Zhang}, {Tanaka}, {Rosero}, {Tan}, {Marvil},
  {Cheng}, {Liu}, {Beltr{\'a}n}, \& {Garay}}]{ZhaY19}
{Zhang}, Y., {Tanaka}, K. E.~I., {Rosero}, V., {et~al.} 2019, \apjl, 886, L4

\end{thebibliography}
%

\end{document}